\documentclass[%
 aip,
 amsmath,amssymb,
 reprint,%
]{revtex4-1}

\usepackage{graphicx}
\usepackage{dcolumn}
\usepackage{bm}

\usepackage[utf8]{inputenc}
\usepackage[T1]{fontenc}
\usepackage{mathptmx}
\usepackage{etoolbox}

\makeatletter
\def\@email#1#2{%
 \endgroup
 \patchcmd{\titleblock@produce}
  {\frontmatter@RRAPformat}
  {\frontmatter@RRAPformat{\produce@RRAP{*#1\href{mailto:#2}{#2}}}\frontmatter@RRAPformat}
  {}{}
}%
\makeatother
\begin{document}
\preprint{AIP/123-QED}

\title[]{Modeling the time-resolved Coulomb explosion imaging of halomethane photodissociation with \textit{ab initio} potential energy curves}
\author{Yijue Ding}
 \affiliation{Department of Physics, Kansas State University, Manhattan, KS 66502, USA}
 \email{yijueding@gmail.com}


\begin{abstract}
We present an effective theoretical model to simulate observables in time-resolved two-fragment Coulomb explosion experiments. The model employs the potential energy curves of the neutral molecule and the doubly charged cation along a predefined reaction coordinate to simulate the photodissociation process followed by Coulomb explosion. We compare our theoretical predictions with pump-probe experiments on iodomethane and bromoiodomethane.
Our theory successfully predicts the two reaction channels in iodomethane photodissociation that lead to $\text{I}(^2\text{P}_{3/2})$ and $\text{I}^*(^2\text{P}_{1/2})$
products, showing excellent agreement with experimental delay-dependent kinetic energy release signals at large pump-probe delays. The theoretical kinetic energy release at small delays depends significantly on the choice of ionic states. By accounting for internal rotation, the kinetic energies of individual fragments in bromoiodomethane align well with experimental results. Furthermore, our theory confirms that two-fragment Coulomb explosion imaging cannot resolve different spin channels in bromoiodomethane photodissociation.
\end{abstract}

\maketitle

\section{Introduction}

Coulomb explosion imaging (CEI) is a powerful technique for retrieving molecular structures due to its equal sensitivity to both light and heavy atoms\cite{vager1989,reviewcei}.
Unlike spectroscopic methods\cite{ftas,atas,trpes}, which are considered indirect imaging techniques, CEI directly retrieves nuclear geometric parameters. In this procedure,
an intense laser pulse rapidly removes multiple electrons from the molecule, inducing strong Coulomb repulsion between the ionic fragments. Detailed molecular structure information is extracted from the final momentum of each fragment\cite{kwon1996,stapelfeldt1998}.

CEI is typically implemented using either velocity map imaging (VMI)\cite{burt2017,burt2018,dingdajun2017,amini2018,corrales2019} or cold target recoil-ion momentum spectrometer (COLTRIM)\cite{vager1989,martin2013,boll2022,endo2020,daniel2023} experimental setups. When used in a static manner, CEI can retrieve complete structural information of organic molecules with more than 10 atoms\cite{boll2022,lam2024}. When implemented in a time-resolved manner during pump-probe experiments, CEI captures snapshots of transient molecular structures, creating the so-called molecular movies. It has been successful in identifying photoisomerization processes\cite{liekhus2015,burt2018}, isolating rare roaming reaction channels\cite{endo2020}, and studying the Jahn-Teller effect\cite{li2021,wang2024}. Due to the low count rate of multi-fragment coincidence detection, time-resolved CEI is usually limited to two-body or three-body fragmentation\cite{cornaggia2009,yatsuhashi2018,li2022b}.

On the theoretical side, models assisting time-resolved CEI in probing photochemical reactions typically involve intricate multi-dimensional molecular dynamics (MD) simulations\cite{ding2024,endo2020,zhou2020}.
Whether performed on-the-fly or on pre-built potential energy surfaces, MD simulations require enormous computational resources.

In this work, we develop a theoretical model that avoids MD simulations to predict observables in time-resolved two-fragment Coulomb explosion experiments. Our model uses \textit{ab initio} potential energy curves of the neutral molecule and the cation to simulate photodissociation followed by Coulomb explosion. The strong-field ionization process that triggers Coulomb explosion is not modeled in detail, and its effect on nuclear motion is ignored.

We apply our model to two representative halomethane species: iodomethane ($\text{CH}_3\text{I}$) and bromoiodomethane ($\text{CH}_2\text{BrI}$). Time-resolved CEI experiments have been conducted on both species using the COLTRIM and VMI setups\cite{corrales2019,daniel2023,burt2017}. In the following sections, we elaborate on our theoretical framework and discuss our results in detail for each halomethane molecule. Finally, we present a brief conclusion summarizing our study.

\section{Iodomethane ($\text{CH}_3\text{I}$)}

The electronic structures are calculated at the multi-reference configuration interaction (MRCI)\cite{mrci1,mrci2} level of theory, using an active space of 6 electrons in 5 orbitals for both the $\text{CH}_3\text{I}$ molecule and the $\text{CH}_3\text{I}^{2+}$ cation. Spin-orbit (SO) coupling effects are treated using the state-interacting approach, in which the SO-coupled Hamiltonian, $\hat{H}_{el}+\hat{H}_{SO}$, is constructed and diagonalized using the eigenfunctions of $\hat{H}_{el}$  as the basis.
Dunning-type correlation-consistent basis sets\cite{dunningbasis} are used in our \textit{ab initio} calculations. Specifically, we choose the cc-pVTZ basis set for hydrogen and carbon atoms and the cc-pVTZ-PP basis set for the iodine atom. In the cc-pVTZ-PP basis, the innermost 28 core electrons of iodine are replaced by a relativistic pseudopotential\cite{pseudopotential,pseudopotential2}. All electronic structure calculations, including those for $\text{CH}_2\text{BrI}$ and $\text{CH}_2\text{BrI}^{2+}$ in the next section, are performed using the MOLPRO quantum chemistry package\cite{molpro,molpro2}.

\begin{figure*}
\includegraphics[width=0.8\textwidth]{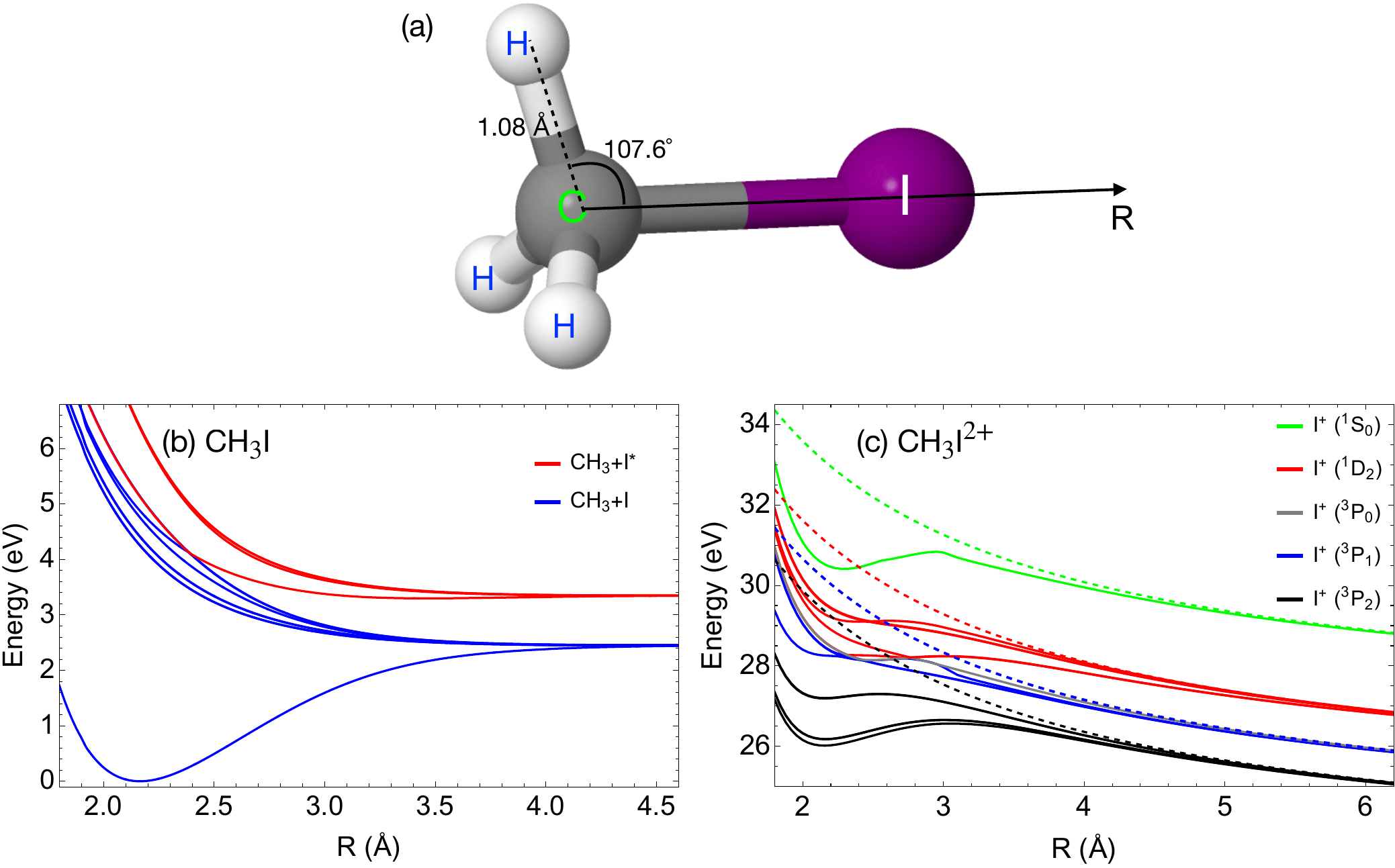}
\caption{(a) Molecular diagram of $\text{CH}_3\text{I}$ in $C_{3v}$ symmetry with the C-I distance $R$ being the reaction coordinate. The remaining DOF are kept at their equilibrium geometry values, as specified in the diagram. (b) Potential energies of $\text{CH}_3\text{I}$ molecule as a function of $R$ for the lowest 12 states that dissociate to $\text{I}$ and $\text{I}^*$ asymptotic thresholds. (c) Potential energies of $\text{CH}_3\text{I}^{2+}$ cation as a function of $R$ for the lowest 15 states that dissociate to different $\text{I}^+$ internal states (indicated by different colors). Energies are given relative to the $\text{CH}_3\text{I}$ ground-state energy at equilibrium. The Coulomb potentials (dashed curves) shifted to the corresponding $\text{I}^+$ thresholds are also shown for comparison.}
\label{fig1}
\end{figure*}   

We calculate 3 singlet states and 3 triplet states of $\text{CH}_3\text{I}$ using the MRCI method, which results in a total of 12 SO-coupled states. These states include the ground state and all A-band excited states of $\text{CH}_3\text{I}$. Regarding the $\text{CH}_3\text{I}^{2+}$ cation, we calculate 7 singlet states and 4 triplet states, which leads to 19 SO-coupled sates in total. Geometry optimizations are initially performed following the ground-state MRCI calculation to obtain the equilibrium geometric parameters of $\text{CH}_3\text{I}$, which are listed in Table \ref{tab1}. We then define a reaction coordinate to construct the potential energy curves with the remaining degrees of freedom (DOF) constrained at the values of the equilibrium geometry. Since $\text{CH}_3\text{I}$ dissociates to $\text{CH}_3+\text{I}$ through UV excitation near 266 nm wavelength, it is natural to choose the C-I distance as the reaction coordinate $R$, as indicated in Fig. \ref{fig1}(a). The potential energies of the 12 states as a function of $R$ are shown in Fig. \ref{fig1}(b). These states dissociate into the $\text{CH}_3+\text{I}$ and $\text{CH}_3+\text{I}^*$ asymptotic thresholds, where $\text{I}(^2\text{P}_{3/2})$ and $\text{I}^*(^2\text{P}_{1/2})$ correspond to two atomic states of iodine with a spin-orbit splitting of 0.9 eV in our calculation. A conical intersection between the 7th and 9th adiabatic states occurs at $R=2.38$ \AA. 15 ionic potential curves along $R$ with asymptotic thresholds corresponding to different internal states of $\text{I}^+$ are shown in Fig. \ref{fig1}(c). These are also compared with the pure Coulomb potential between $\text{CH}_3^+$ and $\text{I}^+$ fragments, assuming the center of charge is located on the center of mass (CM). The real ionic interaction differs significantly from pure Coulomb interaction at small C-I distances due to strong covalent attraction between $\text{CH}_3^+$ and $\text{I}^+$. As the C-I distance increases, the interaction gradually becomes dominated by Coulomb repulsion and is almost purely Coulombic for $R>5$ \AA.

\begin{table}
\caption{Optimized geometric parameters (length in angstrom and angle in degree) of $\text{CH}_3\text{I}$ at equilibrium at the MRCI(6,5)/cc-pVTZ level of theory.}
\begin{ruledtabular}
\begin{tabular}{cc}
 Parameter & Optimized value  \\
\hline
C-I length & 2.16			 		\\
C-H length  & 1.08	 		 \\
I-C-H angle & 107.6	  		\\
H-C-H angle	& 111.3	 		\\
\end{tabular}
\end{ruledtabular}
\label{tab1}
\end{table}

In the time-resolved two-fragment CEI experiment, the pump pulse triggers the photodissociation reaction that results in $\text{CH}_3$ and I radical products, while the probe pulse induce the Coulomb explosion that leads to $\text{CH}_3^+$ and $\text{I}^+$ ionic fragments. Therefore, the total kinetic energy release (KER) can be separated into contributions from these two processes: 
\begin{equation}
\mathrm{KER}=\mathrm{KER_{PD}}+\mathrm{KER_{CE}},
\label{eq1}
\end{equation}
where $\mathrm{KER_{PD}}$ and $\mathrm{KER_{CE}}$ correspond to the contributions from photodissociation and Coulomb explosion, respectively. When CEI is implemented using COLTRIM, the total KER is usually obtained from the final momenta of all ionic fragments measured in coincidence.

In our model, the $\text{CH}_3$ component is treated as a rigid body, and C-I reaction axis goes through the CM of the $\text{CH}_3$ component. Thus, the vibration and rotation of $\text{CH}_3$ component is not excited during photodissociation. In fact, both experiments and MD simulations reveal that over 90\% of the potential energy are redistributed into the translational motion\cite{ding2024}. $\mathrm{KER_{PD}}$ is the sum of CM translational energies of each dissociative component.
Therefore, $\mathrm{KER_{PD}}$ can be written as the potential energy difference between the Franck-Condon (FC) point $R_0$ and the transient geometry $R(t)$ at reaction time $t$: 
\begin{equation}
    \mathrm{KER_{PD}}(t)= V^{(i)}(R_0)-V^{(j)}(R(t)),
\end{equation}
where $i,j$ are indices of the electronic states. $i=j$ denotes adiabatic processes and $i\ne j$ indicates non-adiabatic transition occurs. 
The relation between the transient reaction time $t$ and the reaction coordinate $R$ is given by
\begin{equation}
    t=\int_{R_0}^R\frac{d\Tilde{R}}{\sqrt{2(V^{(i)}(R_0)-V^{(j)}(\Tilde{R}))/\mu}}.
\end{equation}
where $\mu$ is the reduced mass between the two components. 
Since we do not model the strong-field ionization process, $t$ is also treated as the pump-probe delay. Likewise, $\mathrm{KER_{CE}}$ can be written as the ionic potential energy difference between $R$ and the asymptotic threshold: 
\begin{equation}
    \mathrm{KER_{CE}}(t)= V_{ion}^{(i)}(R(t))-V_{ion}^{(j)}(\infty).
\end{equation}
Meanwhile, when we employ pure Coulomb interaction between the ionic fragments and assume that the center of charge is located at the CM, $\mathrm{KER_{CE}}$ can be approximated as
\begin{equation}
\mathrm{KER}_\text{CE}(t)\approx\frac{q_Aq_B}{R_{AB}(t)}
\end{equation}
where $q_A$ and $q_B$ are the charges of A and B fragments (in this case $\text{CH}_3^+$ and $\text{I}^+$), and $R_{AB}$ is the distance between the CMs of A and B fragments. 
\begin{figure}
\includegraphics[width=8.5cm]{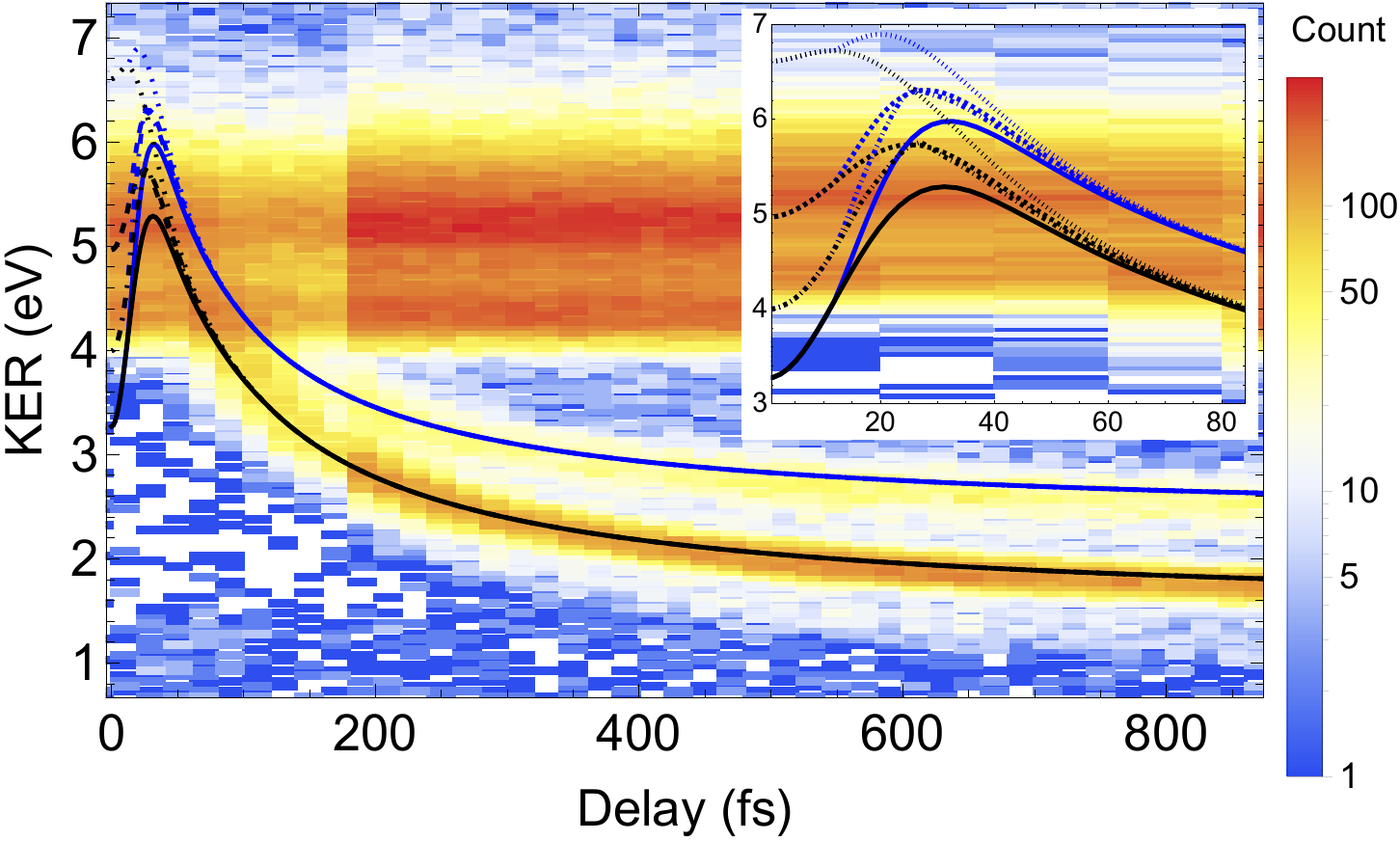}
\caption{The experimental KER signals at different pump-probe delays (adapted from Ref. \onlinecite{daniel2023}) compared with theoretical predictions using \textit{ab initio} potential energy curves. The experimental KER is the sum of $\text{CH}_3^+$ and $\text{I}^+$ kinetic energies measured in coincidence for a UV pump intensity of $1\times10^{13} \text{ W}/\text{cm}^2$.
The photodissociation process starts with the 7th state of $\text{CH}_3\text{I}$ at the FC point and then separate into an adiabatic channel (blue) that remains on the 7th state and a non-adiabatic channel (black) that transitions to the 9th state via a conical intersection at $R=2.38$ \AA. 
The Coulomb explosion is modeled assuming it occurs on the ground state (solid), the 14th state (dashed), and the 15th state (dot-dashed) of $\text{CH}_3\text{I}^{2+}$ cation, as well as using a pure Coulomb potential (dotted). The inset shows the zoom-in of signals at small pump-probe delays. 
}
\label{fig2}
\end{figure}  

To model the photodissociation reaction, the $\text{CH}_3\text{I}$ molecule starts exclusively in the 7th state (the $^3\text{Q}_0$ state in Mulliken notation\cite{mulliken}) at the Franck-Condon (FC) point, as it predominantly (98\%) transitions to this state upon absorbing a UV photon near 260 nm\cite{parker1998,alekseyev20072}. 
The reaction then proceeds along two pathways via the conical intersection at $R=2.38$ \AA. One pathway remains in the 7th state, leading to $\text{CH}_3+\text{I}$ products, while the other undergoes a non-adiabatic transition to the 9th state, resulting in $\text{CH}_3+\text{I}^*$ products.
Typically, the $\text{CH}_3\text{I}$ molecule is most likely ionized to the ground state of the $\text{CH}_3\text{I}^{2+}$ cation through strong-field ionization. Although strong-field ionization can populate excited states of the multi-charged cation, this occurs with a relatively low probability. Therefore, we select the potential energy curves of the ground state, the 14th state, and the 15th state of $\text{CH}_3\text{I}^{2+}$, which correspond to the $\text{I}^+(^3\text{P}_2)$, $\text{I}^+(^1\text{D}_2)$, and $\text{I}^+(^1\text{S}_0)$ thresholds, to model the subsequent Coulomb explosion.

The resulting KER as a function of pump-probe delay compared with the experimental data is shown in Fig. \ref{fig2}. The predicted KER from our theoretical model shows good agreement with the experimental delay-dependent KER signals for large pump-probe delays ($t > 100$ fs), with two major curves corresponding to the two reaction channels leading to I and I* products, respectively. The predictions using different ionic potentials and the pure Coulomb potential are nearly identical at large delays, indicating that the pure Coulomb interaction is a valid approximation in the asymptotic region. At short pump-probe delays ($t < 60$ fs), the dominant delay-independent signals overlap with the delay-dependent ones, showing two major peaks near 4.4 eV and 5.2 eV. The KER values predicted using the ground-state potential of $\text{CH}_3\text{I}^{2+}$ are slightly lower than the experimental values near 0 fs, whereas those obtained using the 14th and 15th ionic states match better with the two peaks observed in the experimental signals. However, our model may underestimate the KER at short delays due to its reduced dimensionality. In fact, MD simulations of the Coulomb explosion on the ground-state potential energy surface of $\text{CH}_3\text{I}^{2+}$ yield  KER of 4.4 eV, which is consistent with the experiment\cite{ding2024}. Meanwhile, the model using the pure Coulomb potential clearly overestimates the KER at short delays.

\section{Bromoiodomethane ($\text{CH}_2\text{BrI}$)}

Similar to our \textit{ab initio} calculations for $\text{CH}_3\text{I}$ and $\text{CH}_3\text{I}^{2+}$, the electronic structures of $\text{CH}_2\text{BrI}$ and $\text{CH}_2\text{BrI}^{2+}$ are also calculated at the MRCI level of theory. The active space is selected to be 12 electrons in 8 orbitals for $\text{CH}_2\text{BrI}$ molecule and 10 electrons in 8 orbitals for $\text{CH}_2\text{BrI}^{2+}$ cation. The basis sets for hydrogen (cc-pVTZ), carbon (cc-pVTZ) and iodine (cc-pVTZ-PP) atoms remain the same as in the $\text{CH}_3\text{I}$ calculation. We also use the cc-pVTZ-PP basis set for the bromine atom, in which the innermost 10 core electrons are replaced by a relativistic pseudopotential.

\begin{table}
\caption{Optimized geometric parameters (length in angstrom and angle in degree) of $\text{CH}_2\text{BrI}$ at equilibrium at the MRCI(12,8)/cc-pVTZ level of theory.}
\begin{ruledtabular}
\begin{tabular}{cc}
 Parameter & Optimized value  \\
\hline
C-I length & 2.16		  	 \\
C-Br length & 1.94	 		\\
C-H length  & 1.08	  		 \\
I-C-Br angle & 114.6	  		\\
H-C-H angle	& 112.3	 		\\
I-C-H angle	& 107.2	  		 \\
Br-C-H angle	& 107.8	 		\\
\end{tabular}
\end{ruledtabular}
\label{tab2}
\end{table}

\begin{figure*}
\includegraphics[width=0.8\textwidth]{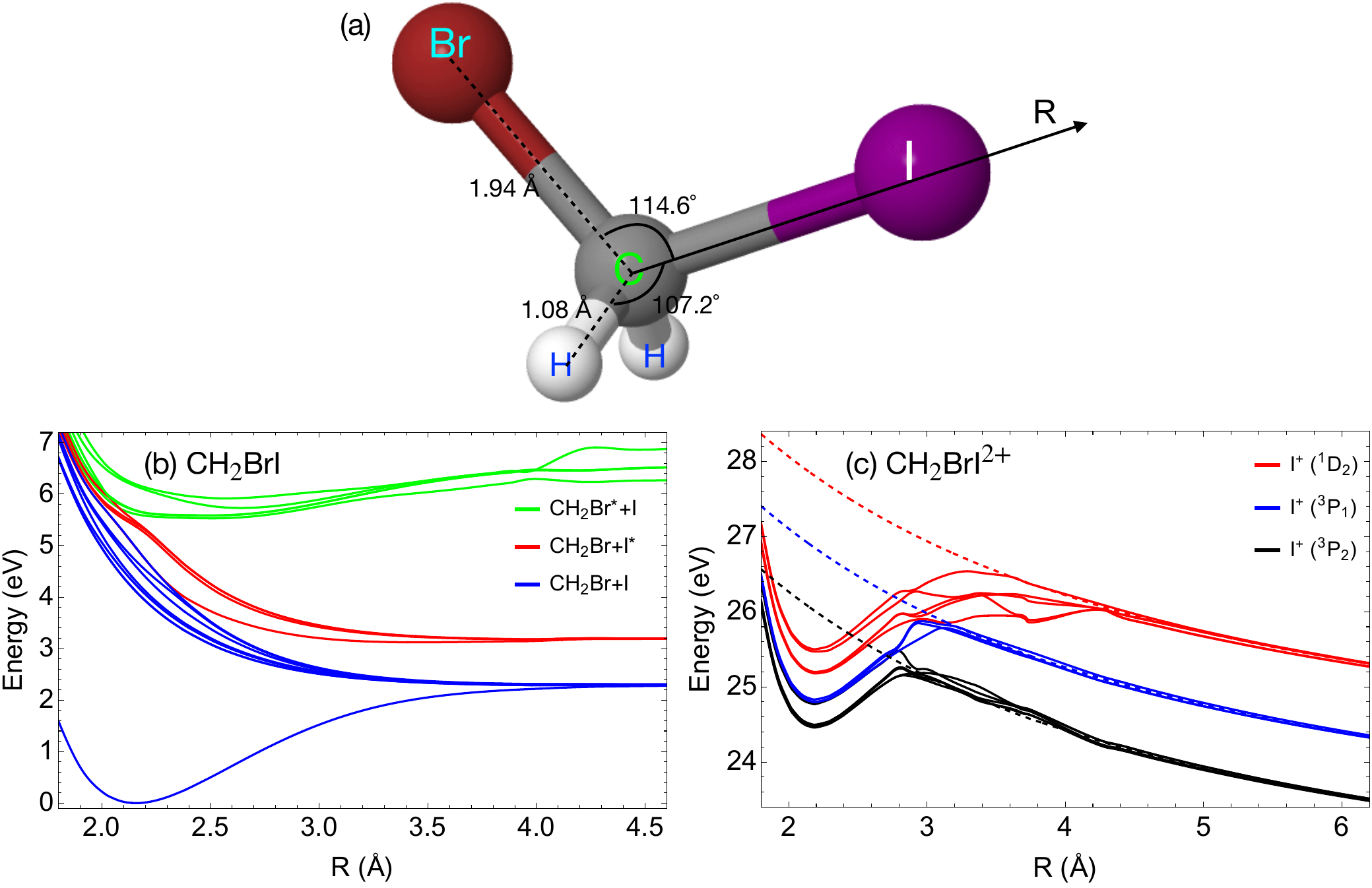}
\caption{(a) Molecular diagram of $\text{CH}_2\text{BrI}$ in $C_s$ symmetry with the C-I distance $R$ being the reaction coordinate. The remaining DOF are kept at the values of the equilibrium geometry, as specified in the diagram. (b) Potential energies of the $\text{CH}_2\text{BrI}$ molecule as a function of $R$ for the lowest 17 states, grouped according to different dissociation thresholds, which are indicated by different colors. (c) Potential energies of the $\text{CH}_2\text{BrI}^{2+}$ cation as a function of $R$ for the lowest 13 states corresponding to different internal states of $\text{I}^+$ ion in the dissociation limit  (indicated by different colors). Energies are given relative to the ground-state energy of $\text{CH}_3\text{BrI}$ at the equilibrium geometry. The pure Coulomb potentials (dashed curves) shifted to the corresponding $\text{I}^+$ thresholds are also shown for comparison.}
\label{fig3}
\end{figure*}

We calculate 5 singlet sates and 4 triplet states for both $\text{CH}_2\text{BrI}$ and $\text{CH}_2\text{BrI}^{2+}$, which 
results in a total of 17 SO coupled states for each species. Geometry optimization is performed following the ground-state MRCI calculation to obtain the equilibrium geometric parameters of $\text{CH}_2\text{BrI}$, which are listed in Table \ref{tab2}. $\text{CH}_2\text{BrI}$ has two major dissociation paths for the A-band excited states, which lead to $\text{CH}_2\text{Br}+\text{I}$ and $\text{CH}_2\text{I}+\text{Br}$ products, respectively. In this work, we focus on the $\text{CH}_2\text{Br}+\text{I}$ reaction path. Therefore, we choose the C-I bond length as the reaction coordinate $R$, and constrain the remaining DOF at the values of the equilibrium geometry, as illustrated in Fig. \ref{fig3}(a). The potential energy curves of the 17 calculated states of $\text{CH}_2\text{BrI}$ along $R$ are shown in Fig. \ref{fig3}(b). These states dissociate into the $\text{CH}_2\text{Br}+\text{I}$, $\text{CH}_2\text{Br}+\text{I}^*$ and $\text{CH}_2\text{Br}^*+\text{I}$ thresholds, where $\text{CH}_2\text{Br}^*$ denotes the excited state of the $\text{CH}_2\text{Br}$ radical. The present calculations do not reveal the spin-orbit splitting of the $\text{CH}_2\text{Br}$ radical. The potential energy curves of 13 ionic states that dissociate into $\text{I}^+(^3\text{P}_2)$, $\text{I}^+(^3\text{P}_1)$, and $\text{I}^+(^1\text{D}_2)$ thresholds are shown in Fig. \ref{fig3}(c). For $R>4\text{ \AA}$, the real ionic potential agrees well with the pure Coulomb potential $1/R_{AB}$, where $R_{AB}$ is the distance between the CMs of $\text{CH}_2\text{Br}^+$ and $\text{I}^+$ fragments. At short distances, the difference between the real ionic potential and the Coulomb approximation is approximately 1 to 2 eV.

\begin{figure*}
\includegraphics[width=0.8\textwidth]{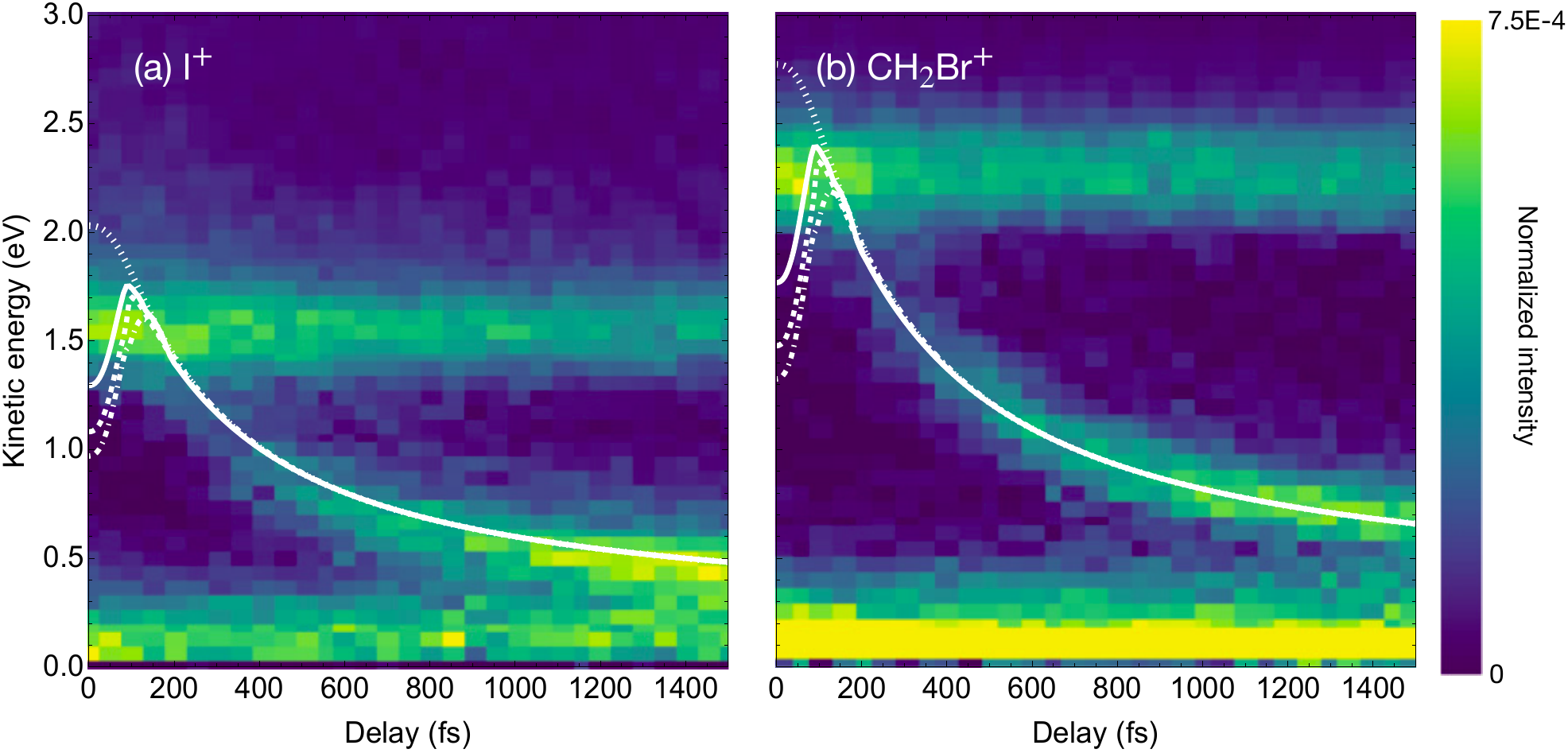}
\caption{The experimental kinetic energy distributions (adapted from Ref. \onlinecite{burt2017}) of $\text{I}^+$ (a) and $\text{CH}_2\text{Br}^+$ (b) at different pump-probe delays compared with theoretical predictions using \textit{ab initio} potential energy curves. The photodissociation is modeled using the 1st excited state of $\text{CH}_2\text{BrI}$. The Coulomb explosion is modeled using the ground state (solid), the 6th state (dashed) and the 13th state (dot-dashed) of the $\text{CH}_2\text{BrI}^{2+}$ cation, as well as the pure Coulomb potential (dotted).}
\label{fig4}
\end{figure*}  

Unlike the $\text{CH}_3\text{I}$ photodissociation, the rotation of $\text{CH}_2\text{Br}$ component is highly excited during the photodissociation of $\text{CH}_2\text{BrI}$\cite{butler1987,burt2017} whereas the $\text{CH}_2\text{Br}$ rotational energy is not measured in the two-fragment CEI experiment. Consequently, only a fraction of the available potential energy is redistributed into $\mathrm{KER_{PD}}$. 
To determine this fraction, we employ the model of a particle colliding with a rigid body with the internal energy released instantaneously. Based on the conservations of momentum, angular momentum, and energy, the equation of motion can be written as
\begin{equation}
\begin{split}
    P_A+P_B=0, \\
    L_A+P_Bh=0,\\
    \frac{P_A^2}{2m_A}+\frac{P_B^2}{2m_B}+\frac{L_A^2}{2I_A}=V^{(i)}(R_0)-V^{(j)}(\infty), 
\end{split}
\label{eq7}
\end{equation}
where A represents the rigid body (in this case the $\text{CH}_2\text{Br}$ component), B represents the particle (the I component). Here, $P_A$ and $P_B$ are the momenta of A and B, respectively, $L_A$ is the angular momentum of A, $I_A$ is the moment of inertia of A, and $h$ is the distance from the CM of A to the reaction axis.

In this model, the asymptotic $\mathrm{KER_{PD}}$ is the sum of CM translational energies of A and B components, which consist of a fraction $\gamma$ of the released potential energy: 
\begin{equation}
    \mathrm{KER_{PD}}(\infty)=\frac{P_A^2}{2m_A}+\frac{P_B^2}{2m_B}=\gamma(V^{(i)}(R_0)-V^{(j)}(\infty))
\end{equation}
Our model reveals $\gamma=0.29$ for $\text{CH}_2\text{BrI}$ photodissociation.
We further approximate the energy fraction $\gamma$ to be a constant during photodissociation, such that the time-dependent $\mathrm{KER_{PD}}$ can be written as
\begin{equation}
    \mathrm{KER_{PD}}(t)\approx \gamma ( V^{(i)}(R_0)-V^{(j)}(R(t)) ).
\end{equation}
Consequently, the relation between the transient reaction time $t$ and the reaction coordinate $R$ is modified to be
\begin{equation}
    t\approx\int_{R_0}^R\frac{d\Tilde{R}}{\sqrt{2\gamma(V^{(i)}(R_0)-V^{(j)}(\Tilde{R}))/\mu}}.
\end{equation}
Since the center of the charge is approximately at the CM of the cation, the Coulomb explosion does not further excite the rotation of the $\text{CH}_2\text{Br}^+$ fragment. 
In the two-component CEI, the kinetic energy of each individual component can be recovered based on the momentum and energy conservation principles, which is given by
\begin{equation}
\begin{split}
   \mathrm{KE_A}=\left(\frac{m_B}{m_A+m_B}\right)\mathrm{KER}, \\
    \mathrm{KE_B}= \left(\frac{m_A}{m_A+m_B}\right)\mathrm{KER}.
\end{split}
\end{equation}

In Fig. \ref{fig4}, we use the 1st excited state of $\text{CH}_2\text{BrI}$ to model the photodissociation and the ground state, the 6th state, and the 13th state of $\text{CH}_2\text{BrI}^{2+}$ to model the Coulomb explosion. At large pump-probe delays, the theoretical kinetic energies of both $\text{I}^+$ (a) and $\text{CH}_2\text{Br}^+$ fragments show good agreement with the CEI experiment implemented using VMI\cite{burt2017}, regardless of the choice of ionic states. At small delays, results using the ground state of $\text{CH}_2\text{BrI}^{2+}$ show better agreement with the experiment than those using other ionic states or the pure Coulomb potential.

\begin{figure*}
\includegraphics[width=0.8\textwidth]{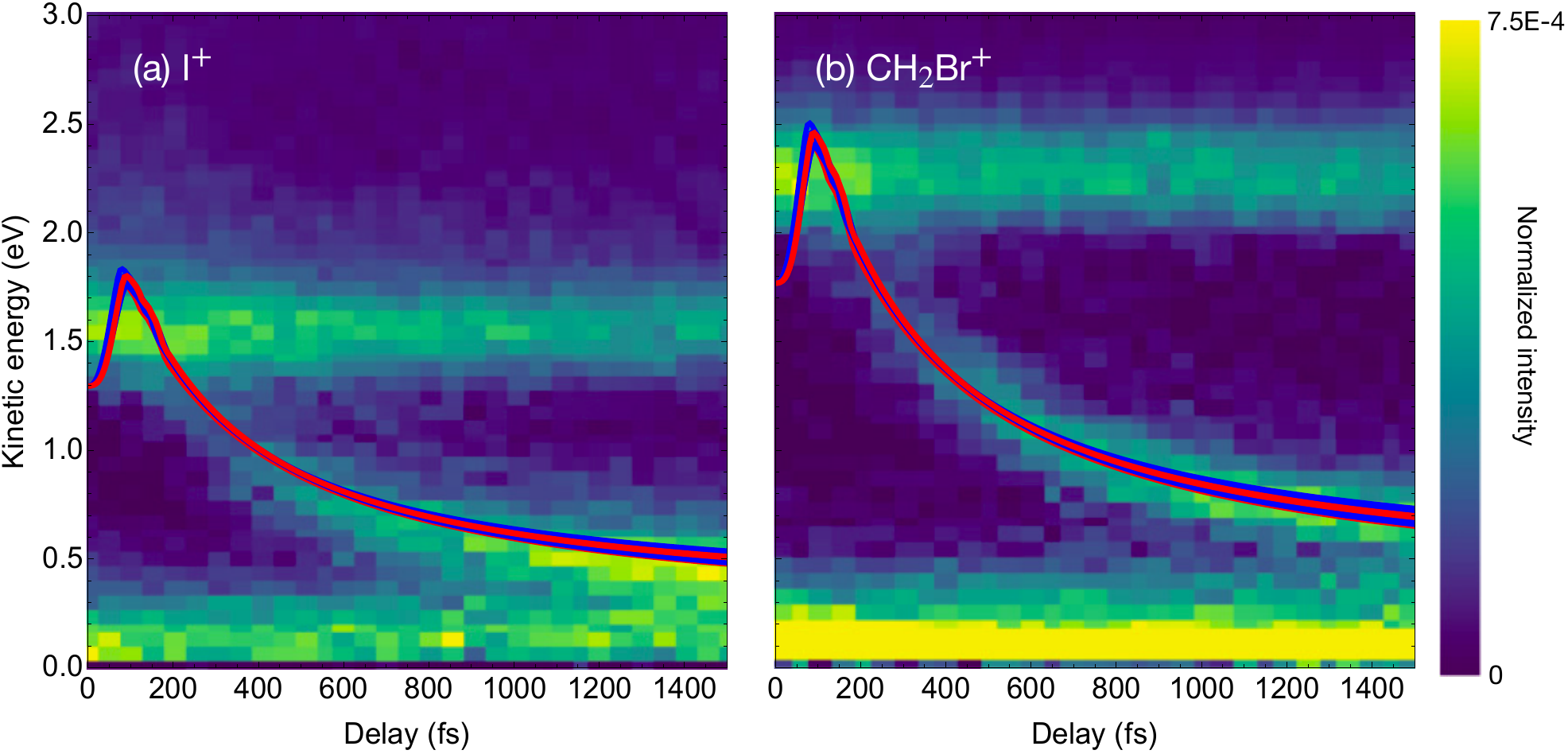}
\caption{The experimental kinetic energy distributions (adapted from Ref. \onlinecite{burt2017}) of  $\text{I}^+$ (a) and $\text{CH}_2\text{Br}^+$ (b) at different pump-probe delays compared with theoretical predictions using \textit{ab initio} potential energy curves. The photodissociation is modeled using all $\text{CH}_2\text{BrI}$ excited states leading to $\text{CH}_2\text{Br}+\text{I}$ (blue) and $\text{CH}_2\text{Br}+\text{I}^*$ (red) dissociation thresholds. The Coulomb explosion is modeled using the ground state of the $\text{CH}_2\text{BrI}^{2+}$ cation. }
\label{fig5}
\end{figure*}

Unlike the time-resolved CEI experiment on $\text{CH}_3\text{I}$, the experiment on $\text{CH}_2\text{BrI}$ cannot resolve different spin channels in photodissociation, which is consistent with our theoretical predictions, as shown in Fig. \ref{fig5}. The theoretical kinetic energy curves for the $\text{CH}_2\text{Br}+\text{I}$ and $\text{CH}_2\text{Br}+\text{I}^*$ adiabatic dissociation channels significantly overlap. This is partly because most of the released potential energy is redistributed into the internal energy of the $\text{CH}_2\text{Br}$ component. Another reason is that, unlike the $\text{CH}_3\text{I}$ molecule, which almost exclusively transitions to the $^3\text{Q}_0$ state, the $\text{CH}_2\text{BrI}$ molecule can transition to multiple excited states upon UV excitation. These excited states near the FC point already differ in potential energy, compensating for the spin-orbit energy splitting in the dissociative thresholds.

\section{Conclusion}

In summary, we have developed a theoretical model to simulate observables in time-resolved two-fragment Coulomb explosion experiments of halomethane species. Our approach employs the potential energy curves of the neutral molecule and the doubly charged cation along a predefined reaction coordinate to simulate photodissociation followed by Coulomb explosion. The potential energy curves are obtained from high-level \textit{ab initio} electronic structure calculations. We have applied our model to the photodissociation and Coulomb explosion of $\text{CH}_3\text{I}$ and $\text{CH}_2\text{BrI}$ molecules. In the $\text{CH}_3\text{I}$ case, the delay-dependent kinetic energy release (KER) shows satisfactory agreement with the COLTRIM experiment\cite{daniel2023} at large pump-probe delays, with two major curves corresponding to the $\text{CH}_3+\text{I}$ and $\text{CH}_3+\text{I}^*$ dissociation channels. At short pump-probe delays, the theoretical KER exhibits noticeable differences depending on the choice of ionic states, with the 14th and 15th states showing better consistency with the CEI experiment. In the $\text{CH}_2\text{BrI}$ case, our model reveals that $\text{CH}_2\text{Br}$ rotation is highly excited during photodissociation, absorbing 71\% of the released potential energy. The individual kinetic energies of the $\text{I}^+$ and $\text{CH}_2\text{Br}^+$ fragments are recovered based on momentum conservation and show good agreement with the VMI measurement\cite{burt2017}, especially when using the ground ionic state potential to model Coulomb explosion. Furthermore, our model explains the experimental observation that different spin channels in $\text{CH}_2\text{BrI}$ photodissociation are not resolved in the kinetic energy signals.

\begin{acknowledgments}
The author thanks Prof. Daniel Rolles for discussions at the early stage of this study. The author thanks Dr. Farzaneh Ziaee for providing the experimental data shown in Fig. 2. The author thanks Kansas State University for providing computing resources that enable the presentation of certain results in this work. 
\end{acknowledgments}

\nocite{*}
\bibliography{aipsamp}

\end{document}